\documentclass[prd,amssymb,amsmath,eqsecnum,showpacs,twocolumn]{revtex4}
\usepackage{epsfig}
\newcommand{\beq}{\begin{equation}}
\newcommand{\eeq}{\end{equation}}
\newcommand{\bqa}{\begin{eqnarray}}
\newcommand{\eqa}{\end{eqnarray}}
\newcommand{\fr}{\frac}

\begin{document}
\title{Shear-free gravitational collapse is strongly censored}
\author{S\'{e}rgio M. C. V. Gon\c{c}alves}
\affiliation{Department of Physics, Yale University, New Haven, Connecticut 06511, USA}
\date{\today}
\begin{abstract}
We consider spherically symmetric spacetimes with matter whose timelike flow is assumed to be shear-free. A number of results on the formation and visibility of spacetime singularities is proven, with the main one being that shear-free collapse {\em cannot} admit locally naked singularities (which implies absence of globally naked singularities). We conjecture that shear is a {\em necessary} condition for the occurrence of locally naked singularities in generic gravitational collapse. 
\end{abstract}
\maketitle

\section{Introduction}

The final state of gravitational collapse is one of the outstanding issues in classical general relativity. One particular aspect of this problem is the role of shear, or absence thereof, in the formation and visibility of singularities. Here, we investigate the formation and local visibility of singularities in spherical shear-free collapse. Our motivation is fourfold: (i) Gravitational thermodynamics considerations suggest that shear affects the endstate of collapse~\cite{penrose79}; (ii) Spherical shear-free spacetimes constitute a vast class, which include cosmological (i.e., non-asymptotically flat) solutions~\cite{krasinski97}, isolated soliton-like objects~\cite{cahill&mcvittie70}, and isolated collapsing objects~\cite{brinisetal00}; (iii) Relativistic kinetic theory tells us that, when the distribution function is isotropic, the solution to the Einstein-Boltzmann equations {\em must} be shear-free~\cite{treciokas&ellis71}; (iv) Shear-free (perfect fluid) solutions constitute a useful model for astrophysical collapse of globular clusters~\cite{glass&mashoon76}.

Within spherical symmetry, studies of certain shear-free collapse models~\cite{glass79,sussman88,brinisetal00,goncalves&jhingan,joshietal03}, as well as Tolman-Bondi (nonvanishing shear) collapse~\cite{joshietal02}, suggest that, provided certain physical reasonability conditions are satisfied, when the matter flow is shear-free resulting spacetime singularities are not naked. In this paper, we consider spherical spacetimes with a general anisotropic fluid, whose timelike flow is assumed to be shear-free; the present analysis includes and generalizes all the spherical shear-free collapse models in the literature thus far.

The main purpose of this paper is threefold: (i) Derive explicit analytical conditions for geodesic completness of the spacetime; (ii) Show that shear-free collapse always leads to a spacetime singularity; (iii) Show that such singularity is locally covered---and thus globally covered~\cite{global}---along {\em all} causal (radial and nonradial) directions. The latter result is the main one in the paper, and may be viewed as a strong cosmic censorship theorem for spherical shear-free spacetimes. We also discuss the interplay between shear and Weyl curvature, and put forward the conjecture that {\em shear is a necessary condition for singularities to be (at least) locally visible in generic gravitational collapse}. By ``generic'', we mean a spacetime of arbitrary geometry, which admits a well-posed regular Cauchy data problem, and whose matter content obeys the weak energy condition. Natural geometrized units, in which $8\pi G=c=1$, are used throughout.

\section{Spherical shear-free spacetimes}

We start with the most general spherically symmetric metric:
\beq
ds^{2}=-e^{2\Psi}dt^{2}+e^{2\Phi}dr^{2}+R^{2}(d\theta^{2}+\sin^{2}\theta d\varphi^{2}), \label{metr}
\eeq
where $t\in{\mathbb R}$, $r\in{\mathbb R}_{0}^{+}$, $\theta\in[0,\pi]$, $\varphi\in[0,2\pi)$, and $\Psi$, $\Phi$, and $R$ are functions of $t,r$ alone. For the matter content we take the stress-energy tensor:
\beq
T^{\mu}_{\nu}=\mbox{diag}(-\rho,\Pi,\Gamma,\Gamma), \label{stress}
\eeq
where $\Pi$ and $\Gamma$ are the principal pressures, parallel and orthogonal to the timelike fluid flow $u^{\mu}=e^{-\Psi}\delta^{\mu}_{t}$, respectively. The field equations may be written as:
\bqa
R_{,tr}&=&\Phi_{,t}R_{,r}+\Psi_{,r}R_{,t}, \label{e1} \\
M_{,r}&=&\fr{1}{2}\rho R^{2}R_{,r}, \label{e2} \\
M_{,t}&=&-\fr{1}{2}\Pi R^{2}R_{,t}, \label{e3} \\
\Pi_{,r}&=&2(\Gamma-\Pi)\fr{R_{,r}}{R}-(\rho+\Pi)\Psi_{,r}, \label{e4} \\
M(t,r)&:=&\fr{R}{2}(1-g^{\mu\nu}R_{,\mu}R_{,\nu}), \label{msm}
\eqa
where the last equation is the Misner-Sharp mass~\cite{misner&sharp64}. Solutions with the stress-energy tensor (\ref{stress}) are fully characterized by the kinematical properties of the 4-velocity $u^{\mu}$: vorticity $\omega_{\mu\nu}$, acceleration $a_{\mu}$, expansion $\Theta$, and shear $\sigma_{\mu\nu}$:
\bqa
\omega_{\mu\nu}&=&0, \\
a_{\mu}&=&\Psi_{,r}\delta^{r}_{\mu}, \\
\Theta&=&e^{-\Psi}(\Phi_{,t}+2R_{,t}R^{-1}), \label{exp} \\
\sigma^{\mu}_{\nu}&=&\mbox{diag}(0,-2\sigma,\sigma,\sigma),
\eqa
where the vorticity vanishes because of spherical symmetry, and 
\beq
\sigma\equiv\fr{1}{3}e^{-\Psi}(R_{,t}R^{-1}-\Phi_{,t}).
\eeq
Shear-free motion is given by $R_{,t}R^{-1}=\Phi_{,t}$, which integrates to
\beq
e^{\Phi}=f(r)R, \label{sfree}
\eeq
where $f(r)$ is an arbitrary positive-definite real-valued function, which we set as $f(r)=1/r$ without loss of generality. Equation (\ref{e1}) then integrates to
\beq
e^{2\Psi}=\fr{(R_{,t})^{2}}{R^{2}H^{2}(t)},
\eeq 
where $H(t)$ is (one third of) the expansion of the flow lines, as can be seen from Eq. (\ref{exp}). For the case of perfect fluids ($\Pi=\Gamma$), the remaining field equations lead to a single second-order nonlinear partial differential equation~\cite{wyman46qvist48}:
\beq
F_{,xx}+J(x)F^{2}=0, \label{sfs}
\eeq
where $x:=r^{2}$, $F(t,r):=\sqrt{x}/R$, and $J(x)$ is an arbitrary function. Equation (\ref{sfs}) fully determines the solution to spherical shear-free perfect fluid collapse. For $J_{,x}=0$, the PDE can be solved in terms of Weierstrass $\wp$-functions~\cite{abramovitz&stegun65}, and physically it corresponds to the so-called synchronous evolution, wherein all shells collapse to vanishing proper area in the same amount of proper time~\cite{brinisetal00}. For $J_{,x}\neq0$, Eq. (\ref{sfs}) cannot in general be solved in closed form, and very few cases admit analytical solutions~\cite{ince26}. In this paper we shall consider the general case ($\Pi\neq\Gamma$), noting, however, that all of the subsequent results also hold for the perfect fluid case.

We now impose the following regularity conditions on the metric:
\bqa
&&R(t,0)=R_{,t}(t,0)=0, \label{c1} \\
&&\lim_{r\rightarrow0} r\fr{R_{,r}}{R}=1, \label{c2} \\
&&g_{\mu\nu}\in C^{1}({\mathbb R}) \;\;\; \mbox{and} \;\;\; |g_{\mu\nu,\alpha}|_{r=0}<\infty. \label{c3}
\eqa
We note that, although $C^{2}$ differentiability is required for non-singular Riemann curvature, we shall not impose it (other than on the initial Cauchy data, as detailed below), since one would like to allow for curvature singularities, which arise dynamically from the gravitational collapse of regular Cauchy data; in addition, local flatness and geodesic motion only require $g_{\mu\nu}\in C^{1}({\mathbb R})$, since only the Christoffel connection enters the geodesic equation.

In spherical symmetry there are no gravitational degrees of freedom, and the metric is therefore uniquely determined from Cauchy data on a spacelike hypersurface $\Sigma_{t_{0}}=\{x^{\mu}: x^{0}=t_{0}=\mbox{const.}\}$. To ensure that any putative singularities arise dynamically from {\em regular} initial data, we impose the following conditions on it:
\bqa
&&g_{\mu\nu}(t_{0},r)\in C^{2}({\mathbb R}), \label{cc1} \\
&&2M(t_{0},r)<R(t_{0},r). \label{cc2}
\eqa
Condition (\ref{cc1}) ensures smooth Riemann curvature on $\Sigma_{t_{0}}$, and (\ref{cc2}) guarantees the initial slice is free of trapped surfaces.

For the collapse of isolated bodies, the metric (\ref{metr}) must be matched to a Schwarzschild exterior at some finite coordinate radius $r_{\rm c}$, via the Darmois-Israel junctions conditions, which impose continuity of the metric and extrinsic curvature across the junction surface. Since we shall be interested in the $R=0$ singularity arising from gravitational collapse of {\em any} spherical shear-free spacetime---not necessarily asymptotically flat---we shall not discuss the details of the matching (see, e.g.,~\cite{herreraetal02}), which fixes the function $H(t)$ for a given interior solution $\{\Psi,\Phi,R,\rho,\Pi,\Gamma\}$. For collapsing solutions $R_{,t}<0$ and $H(t)<0$, and the particular functional form of $H(t)$ is qualitatively irrelevant for the endstate of collapse.

Before proceeding, we compute the Brown-York quasilocal mass~\cite{brown&york93}, which provides a useful tool to analyze the role of ``energy'' in the formation of curvature singularities. Consider a spacelike hypersurface $\Sigma_{t}$ with two-boundary $\partial\Sigma_{t}$; the quasilocal mass $E$ on $\Sigma_{t}$ associated with the three-volume defined by $\partial\Sigma_{t}$ is {\em the value of the Hamiltonian that generates unit time translations orthogonal to the boundary $\partial\Sigma_{t}$}~\cite{brown&york93}. In terms of the induced two-metric $\gamma_{ab}$ on $\partial\Sigma_{t}$, it is given by the proper surface integral~\footnote{Here, and in Eq. (\ref{bym}), we use $G=c=1$ instead of $8\pi G=c=1$, so as to reproduce exactly the formula given by Brown and York in~\cite{brown&york93}.}
\beq
E:=\fr{1}{8\pi}\int_{\partial\Sigma_{t}} d^{2}x  \sqrt{\gamma} (K-K_{0}), \label{eby}
\eeq
where $K$ is the trace of the extrinsic curvature of $\partial\Sigma_{t}$ as imbedded in $\Sigma_{t}$, and $K_{0}$ is the corresponding quantity for the {\em same} surface as embedded in a three-dimensional slice of flat spacetime (such embedding is assumed to be possible). This last term corresponds to a normalization of the ``zero of energy'' with respect to flat space. 

In our case, the Riemannian metric on $\Sigma_{t}$ is
\beq
h_{ij}dx^{i}dx^{j}=e^{2\Phi}dr^{2}+R^{2}d\theta^{2}+R^{2}\sin^{2}\theta d\varphi^{2}. 
\eeq
For the two-boundary $\partial\Sigma_{t}$ we take the surface ${\mathcal S}$, parametrically defined by $\Xi(x^{\mu})=r-r_{0}=0$, which has the induced Riemannian two-metric:
\beq
\gamma_{ab}dw^{a}dw^{b}=R^{2}(t,r_{0})d\theta^{2}+R^{2}(t,r_{0})\sin^{2}\theta d\varphi^{2}.
\eeq
The extrinsic curvature of ${\mathcal S}$ as embedded in $\Sigma_{t}$ is
\beq
K_{ab}=n_{i}\left(\fr{\partial^{2} x^{i}}{\partial w^{a}\partial w^{b}}+\,^{(3)}\Gamma^{i}_{jk}\fr{\partial x^{j}}{\partial w^{a}}\fr{\partial x^{k}}{\partial w^{b}}\right),
\eeq
where
\beq
n_{i}=\fr{^{(3)}\nabla_{i}\Xi}{\sqrt{h^{ij}\,^{(3)}\nabla_{i}\Xi\,^{(3)}\nabla_{j}\Xi}}=e^{\Phi}\delta_{i}^{r}
\eeq
is the unit normal to ${\mathcal S}$. The only non-vanishing components of $K_{ab}$  are
\beq
K_{\varphi\varphi}=\sin^{2}\theta K_{\theta\theta}=-e^{-\Phi}RR_{,r},
\eeq
whence
\beq
K=K_{ab}\gamma^{ab}=-2e^{-\Phi}\fr{R_{,r}}{R}.
\eeq
For the Euclidean reference surface, we take the metric
\beq
E_{ij}dx^{i}dx^{j}=dr^{2}+r^{2}d\theta^{2}+r^{2}\sin^{2}\theta d\varphi^{2}.
\eeq
A trivial calculation gives $K_{0}=-2/r$, and the Brown-York mass is then
\bqa
E&=&\fr{1}{8\pi}\int_{0}^{2\pi} d\varphi \int_{0}^{\pi} d\theta \, r^{2}\sin\theta\left(-2e^{-\Phi}\fr{R_{,r}}{R}+\fr{2}{r}\right) \nonumber \\
&=&r\left(1-e^{-\Phi}r\fr{R_{,r}}{R}\right), \label{bym}
\eqa
where all quantities are evaluated at $r=r_{0}$ on $\Sigma_{t}$. This quasilocal mass agrees with the ADM and Bondi-Sachs mass in the appropriate limits, and a ``positive-mass'' theorem for it has recently been proven~\cite{liu&yau03}. To make contact with the Schwarzschild spacetime (with mass parameter $m$), set $R=r$ and $\Phi=-\ln\sqrt{1-2m/r}$ in Eq. (\ref{bym}), whereby Eq. (6.14) in~\cite{brown&york93} is recovered. This shows that $E$ asymptotes $m$ (and thus $M$) in the weak-field limit, $m/r\ll1$.

\section{Causal geodesic completeness}

A spacetime is said to be causally g-complete if every non-spacelike geodesic $\gamma(\lambda)$ can be extended to arbitrary values of its affine parameter~\cite{hawking&ellis73}. One must note that g-completeness does {\em not} imply that the spacetime is singularity-free, since non-geodesic (but causal) inextendible curves {\em can} exist in g-complete spacetimes~\cite{geroch68}. A spacetime is singularity-free if {\em all} $C^{1}$ curves, geodesic or otherwise, are extendible, in which case it is said to be b-complete; b-completeness implies g-completeness, but the converse is not true. Evidently, then, failure of g-completeness is sufficient to show that the spacetime contains a singularity. From elementary ODE theory, a necessary and sufficient condition for g-completeness is the boundedness of {\em all} of the first derivatives of the coordinates with respect to the affine parameter~\cite{arnold73}. Since we are interested in showing that shear-free collapse always leads to a spacetime singularity, it suffices to take a particular family of causal geodesics and show that it is inextendible, i.e., that at least one of the functions $dx^{\mu}/d\lambda$ grows unbounded for finite $\lambda$.

We start with a generic geodesic, $\gamma(\lambda)$, defined by the integral curves of the vector field
\beq
\xi^{\mu}=\fr{dx^{\mu}}{d\lambda}=(\dot{t},\dot{r},\dot{\theta},\dot{\varphi}),
\eeq
where $\lambda$ is an affine parameter along $\gamma(\lambda)$, and the overdot denotes total differentiation with respect to $\lambda$. Normalizing $\xi^{\mu}\xi_{\mu}=-\epsilon$, we have
\beq
\dot{t}^{2}=e^{-2\Psi}\left[e^{2\Phi}\dot{r}^{2}+\fr{L^{2}}{R^{2}}+\epsilon\right], \label{ge0}
\eeq
where $L$ is the total conserved angular momentum, given by $L^{2}=L_{(\theta)}^{2}+L_{(\varphi)}^{2}$, where
\beq
L_{(\theta)}=\delta^{\mu}_{\theta}\xi_{\mu}=\dot{\theta}R^{2}, \;\;
L_{(\varphi)}=\delta^{\mu}_{\varphi}\xi_{\mu}=\dot{\varphi}R^{2}\sin^{2}\theta. 
\eeq
We remark that, without loss of generality, one can always take the polar coordinates to be such that $\gamma(\lambda)$ moves along the $\theta=\pi/2$ hypersurface, whence $L=\dot{\varphi}R^{2}$; obviously, this does not change Eq. (\ref{ge0}), nor the geodesic equation:
\bqa
2\ddot{t}&+&\dot{t}^{2}(\Psi_{,t}+\Phi_{,t})+2\dot{t}\dot{r}\Psi_{,r}+\fr{L^{2}}{R^{2}}e^{-2\Psi}\left(\fr{R_{,t}}{R}-\Phi_{,t}\right) \nonumber \\
&-&\epsilon e^{-2\Psi}\Phi_{,t}=0, \label{ge1} \\
2\ddot{r}&+&\dot{r}^{2}(\Psi_{,r}+\Phi_{,r})+2\dot{t}\dot{r}\Phi_{,t}-\fr{L^{2}}{R^{2}}e^{-2\Phi}\left(\fr{R_{,r}}{R}-\Psi_{,r}\right) \nonumber \\
&-&\epsilon e^{-2\Phi}\Psi_{,r}=0. \label{ge2}
\eqa
Equations (\ref{ge0}), (\ref{ge1})--(\ref{ge2}) form a complete set for the three variables $\{\dot{t},\dot{r},\dot{\varphi}\}$.

We now specialize to the case of future-oriented ingoing radial null geodesics (IRNG): 
\bqa
&&\dot{t}>0, \;\; \dot{r}<0, \;\; \dot{\theta}=\dot{\varphi}=\epsilon=0, \label{ge0n} \\
&&2\ddot{t}+\dot{t}^{2}(\Psi_{,t}+\Phi_{,t})+2\dot{t}\dot{r}\Psi_{,r}=0, \label{ge1n} \\
&&2\ddot{r}+\dot{r}^{2}(\Psi_{,r}+\Phi_{,r})+2\dot{t}\dot{r}\Phi_{,t}=0. \label{ge2n}
\eqa
The null condition $\epsilon=0$ reduces these two equations to a single first-order ODE for $\dot{t}$:
\beq
\fr{1}{\dot{t}}\fr{d\dot{t}}{d\lambda}=-\fr{1}{2}\left[\dot{\Psi}+\dot{\Phi}+\dot{t}(\Phi_{,t}-\Psi_{,t})\right], \label{orng}
\eeq
which integrates to
\beq
\dot{t}=\beta(r)e^{-\Phi}, \label{geot}
\eeq
where $\beta(r): {\mathbb R}_{0}^{+}\mapsto {\mathbb R}^{+}\backslash\{0\}$ is an arbitrary integration function. We can now put forward the following results:

Proposition 1. {\em A spherically symmetric spacetime with metric (\ref{metr}) is g-complete along geodesics defined by (\ref{ge0n})-(\ref{ge2n}), if the following holds:}
\beq
\exists C\in{\mathbb R}^{+}\backslash\{0\}: \; \Phi(t,r)>\ln C+f(r), \;\; \forall t\in{\mathbb R}, \, \forall r\in{\mathbb R}_{0}^{+}. \nonumber
\eeq

Proof. The condition above implies that $\dot{t}$ remains finite and bounded away from zero. Since $|\dot{r}|=\dot{t}e^{\Psi-\Phi}$, and $\{\Psi,\Phi\} \in C^{1}({\mathbb R})$ with finite first derivatives [cf. Eq. (\ref{c3})], it follows that $\dot{r}$ is also finite and bounded away from zero. The first derivative of all the coordinates with respect to the affine parameter is always finite along IRNG, and therefore the spacetime is g-complete along IRNG. $\Box$

Proposition 2. {\em Consider a spacetime with metric (\ref{metr}) which obeys the Einstein equations for the stress-energy tensor (\ref{stress}). Assume that some portion of the spacetime undergoes complete gravitational collapse, i.e., $\exists\, t_{\rm s}\in{\mathbb R}^{+}\backslash\{0\}: R(t_{\rm s},r)=0$. If the shear-free condition (\ref{sfree}) is satisfied, such collapsing solutions are g-incomplete.}

Proof. This follows from the shear-free condition (\ref{sfree}): for $r>0$, $R\searrow0\Rightarrow\Phi\searrow-\infty$, which implies---by Eq. (\ref{geot})---that $\dot{t}\nearrow+\infty$. Consider now $0<\varepsilon\ll1$, whence $R(t_{\rm s},\varepsilon)=\varepsilon e^{\Phi(t_{\rm s},\varepsilon)}=0\Rightarrow \Phi(t_{\rm s},\varepsilon)\searrow-\infty$. By continuity, $\lim_{\varepsilon\rightarrow0} \Phi(t_{\rm s},\varepsilon)=-\infty$, and thus $\Phi(t_{\rm s},r)\searrow-\infty, \forall\, r\in {\mathbb R}_{0}^{+}$. This means that IRNG are future-inextendible, and thus the spacetime is g-incomplete. We remark that, by standard extendability results for causal curves, one can show that future-directed timelike geodesics are also incomplete at $R=0$. $\Box$

Proposition 3. {\em A necessary and sufficient condition for the spacetime to contain a singularity is that the Brown-York quasilocal mass diverges negatively, $E\searrow-\infty$.}

Proof. For $r>0$, by Eq. (\ref{bym}), $E\searrow-\infty$ implies $\Phi\searrow-\infty$, and hence $R\searrow0$ by Eq. (\ref{sfree}). The converse is also true: if $R=0$ with $r>0$, then $\Phi\searrow-\infty$ and thus $E\searrow-\infty$. For $r=0$, Eq. (\ref{bym}) and the regularity condition (\ref{c2}) imply
\beq
\lim_{r\rightarrow0} E=-e^{-\Phi}r^{3}. \label{bym0}
\eeq
If $E\searrow-\infty$ then $\Phi\searrow-\infty$, whence $R\searrow0$. To prove the converse, we first note that regularity at the center [Eq. (\ref{c2})] together with Eq. (\ref{sfree}) imply that $\lim_{r\rightarrow0} \Phi_{,r}=0$. As shown previously (cf. proof of Prop. 2), $\lim_{r\rightarrow0} \Phi=-\infty$, which means that the limit (\ref{bym0}) is a $0/0$-undeterminacy. Successive application of l'H\^{o}pital's rule, together with $\lim_{r\rightarrow0} \Phi_{,r}=0$ yields $E\searrow-\infty$. This completes the proof. $\Box$

\section{Local visibility of the singularity}

Given that a spacetime singularity forms at $R(t_{\rm s},r)=0$, is it (at least) locally visible? Local visibility requires one to show the
existence of non-spacelike future-directed outgoing geodesics---say, ORNG---with their past endpoint at the
singularity. Geometrically, this is equivalent to requiring that the area radius increases along such geodesics, i.e., $(dR/dr)_{\rm ORNG}>0$.
Along ORNG we have
\beq
\fr{dR}{dr}=R_{,r}+R_{,t}\left(\fr{dt}{dr}\right)_{\rm ORNG}=R_{,r}+e^{\Phi-\Psi}R_{,t}.
\eeq
Using the standard procedure \cite{joshi&dwivedi93}, we introduce auxiliary variables $u$, $X$:
\beq
u\equiv r^{\alpha}\;,\;\alpha>0; \;\;\;\; X\equiv\fr{R}{u}. \label{Xdef}
\eeq
In the limit of approach to the singularity we have
\bqa
X_{0}&\equiv&\lim_{R\rightarrow0,u\rightarrow0}\fr{R}{u}=
\lim_{R\rightarrow0,u\rightarrow0}\fr{dR}{du} \nonumber \\
&=&\lim_{R\rightarrow0,r\rightarrow0}\fr{1}{\alpha
r^{\alpha-1}}\fr{dR}{dr} \nonumber \\
&=&\lim_{R\rightarrow0,r\rightarrow0}\fr{1}{\alpha
r^{\alpha-1}}\left(R_{,r}+e^{\Phi-\Psi}R_{,t}\right) \nonumber \\
&=&\lim_{r\rightarrow0} \fr{e^{\Phi}}{\alpha r^{\alpha-2}}\left(\Phi_{,r}+\fr{e^{\Phi}}{H(t)}+\fr{1}{r}\right). \label{lv}
\eqa
$X_{0}$ is the value of the tangent to the geodesic on the $u-X$
plane, at the singularity. If $X_{0}$ is real and positive-definite, the singularity
is at least locally naked, and covered otherwise.

Proposition 4. {\em The $R(t_{\rm s},0)=0$ singularity is also a shell-crossing singularity.}

Proof. Shell-crossing singularities occur when $R_{,r}=0$, and signal the overlapping of infinitesimally close spherical shells on a given spacelike slice $\Sigma_{t}$. Such singularities also occur in Newtonian collapse, and, as such, are not intrinsic general-relativistic phenomena; in addition, for $R>0$, they are gravitationally weak, and may be transformed away by a $C^{1}$ coordinate change~\cite{shellx}. To distinguish shell-crossings from shell-focusing singularities, the condition $R>0$ is usually implied in the definition of the former; however, one can have $R=R_{,r}=0$, which is the case addressed here. Regularity at the center [cf. Eq. (\ref{c2})] together with the shear-free condition (\ref{sfree}) [with the gauge choice $f(r)=1/r$] imply $\lim_{r\rightarrow0} R_{,r}=e^{\Phi(t_{\rm s},0)}$. As shown previously, $\lim_{r\rightarrow0} \Phi(t_{\rm s},r)=-\infty$, and thus $\lim_{r\rightarrow0} R_{,r}(t_{\rm s},r)=0$. $\Box$

Proposition 5. {\em The $R(t_{\rm s},0)=0$ singularity in shear-free collapse is covered.}

Proof. The function $H(t)$ is monotonically decreasing for spacetimes undergoing complete gravitational collapse [for spacetimes with a Schwarzschild exterior matched to the interior solution at $r=r_{\rm c}$, $H(t)$ diverges negatively at $t=t_{\rm s}(r_{\rm c})$, when the ``last'' matter shell collapses, signalling infinite convergence of the fluid flow], and regularity at the center requires $\lim_{r\rightarrow0} \Phi_{,r}=0$. From Eq. (\ref{lv}) we have then
\beq
X_{0}=\lim_{r\rightarrow0} \fr{e^{\Phi(t_{\rm s},r)}}{\alpha r^{\alpha-1}}.
\eeq
The requirement $\Phi_{,r}\rightarrow0$ as $r\searrow0$ implies that one cannot have $e^{\Phi}\propto r^{\alpha-1}$. This, in turn, implies that, for $\alpha\in(0,1]$ we have $X_{0}=0$, and the singularity is thus covered. For $\alpha>1$ there are two distinct possibilities (with the necessary assumption that $e^{\Phi}\not\propto r^{\alpha-1}$): (i) $e^{\Phi}>r^{\alpha-1}$, or (ii) $e^{\Phi}<r^{\alpha-1}$. Possibility (i) implies that $X_{0}\nearrow+\infty$, and (ii) yields $X_{0}=0$; that is, for $\alpha>1$, $X_{0}$ cannot be positive-definite, and the singularity is covered. $\Box$

Proposition 6. {\em If the weak energy condition holds, $\Pi>0$, and $\rho<1/R^{2}$, all noncentral singularities $R(t_{\rm s},r>0)=0$ are covered.}

Proof. The sufficient condition for local visibility outlined at the beginning of this section only works in the $r\searrow0$ limit, so one must resort to other methods to study the local visibility of noncentral ($r>0$) singularities. Noncentral shells will be covered if they become trapped before they become singular. The apparent horizon in spherical symmetry is given by $R(t_{\rm ah},r)=2M(t_{\rm ah},r)$, which implicitly defines the curve $t_{\rm ah}(r)$. On the $t-R$ plane, the condition for local coveredness is then:
\beq
\fr{dt_{\rm ah}}{dR}=\fr{dt_{\rm ah}}{dr}/\left(\fr{dR}{dr}\right)_{\rm ah}=\fr{t'_{\rm ah}}{R_{,r}+R_{,t}t'_{\rm ah}}>0. \label{covered}
\eeq
where $t'_{\rm ah}=dt_{\rm ah}/dr$. Now, since $R_{,t}<0$, the above condition reads
\bqa
t'_{\rm ah}&>&0, \label{ccv1} \\
R_{,r}+R_{,t}t'_{\rm ah}&\geq&0. \label{ccv2}
\eqa
Taking the total $r$-derivative of $R(t_{\rm ah},r)=2M(t_{\rm ah},r)$ along $t_{\rm ah}(r)$ one obtains
\beq
t'_{\rm ah}=\fr{2M_{,r}-R_{,r}}{R_{,t}-2M_{,t}}, \label{tahp}
\eeq
where all quantities are evaluated along $t_{\rm ah}(r)$. The condition $\Pi>0$ together with Eq. (\ref{e3}) implies $M_{,t}>0$, and $\rho<1/R^{2}$ together with Eq. (\ref{e2}) implies $2M_{,r}<R_{,r}$. This means that both the numerator and denominator in Eq. (\ref{tahp}) are negative, whereby condition (\ref{ccv1}) is satisfied. Now, since $R_{,t}<0$ and $M_{,t}>0$, condition (\ref{ccv2}) is equivalent to $M_{,r}R_{,t}-M_{,t}R_{,r}\leq0$. Using Eqs. (\ref{e2})--(\ref{e3}), this reads
\beq
R^{2}R_{,t}R_{,r}(\rho+\Pi)\leq0,
\eeq
which holds provided the weak energy condition holds: $T_{\mu\nu}u^{\mu}u^{\nu}\geq0\Rightarrow\rho+\Pi\geq0$. This completes the proof. $\Box$ 

Proposition 7. {\em All radially null-covered $R=0$ singularities are covered along {\em all} causal directions.}

Proof. Radially null-covered $R=0$ singularities satisfy $t_{\rm ah}(r)<t_{*}(r)$, where $t_{*}(r)$ is the solution for ORNG on the $t-r$ plane, and is given by the first-order ODE:
\beq
\fr{dt_{*}}{dr}={\mathcal F}(t,r)\equiv e^{\Phi-\Psi},
\eeq
with the initial datum $t_{*}(0)=t_{\rm s}(0).$ Future-directed nonradial geodesics are given by
\beq
\fr{d\hat{t}}{dr}=e^{\Phi-\Psi}\sqrt{1+\fr{e^{-2\Phi}}{\dot{r}^{2}}\left(\fr{L^{2}}{R^{2}}+\epsilon\right)}>{\mathcal F}(t,r),
\eeq
with $\hat{t}(0)=t_{\rm s}(0)$. Since $\hat{t(0)}=t_{*}(0)$, and $(d\hat{t}/dr)>(dt_{*}/dr)>0$, by a straightforward application of the Gronwall lemma it follows that 
\beq
\hat{t}(r)\geq t_{*}(r), \;\; \forall \, r\in {\mathbb R}_{0}^{+},
\eeq
where the inequality saturates only for $r=0$. Since $t_{\rm ah}(r)<t_{*}(r)$, we also have $t_{\rm ah}(r)<\hat{t}(r)$, which means that the singularity is covered along {\em all} causal (radial and nonradial) directions. $\Box$

\section{Concluding remarks}

We have shown that, provided certain plausible regularity and physical reasonability conditions hold, spherical shear-free collapse results in a locally covered singularity, which must therefore be spacelike. This may thus be viewed as a strong cosmic censorship result for the class of spacetimes considered herein. In terms of matter content, such class includes virtually all known classical descriptions of matter---anisotropic fluids and specializations thereof (e.g., perfect fluids, dust), Maxwell fields, and scalar fields---and the constraints imposed on it (needed {\em only} for the particular case of noncentral shells; cf. Prop. 6) are arguably soft.

What is the {\em gravitational} nature of the spacelike singularity in shear-free collapse? In typical collapse situations (with nonvanishing shear) leading to a curvature singularity, the Weyl curvature diverges faster than the Ricci, near the singularity~\cite{weylricci}. Now, shear-free collapse altogether {\em forbids} the influence of the Weyl tensor on the evolution of nonspacelike congruences, since only the propagation equation for shear contains a Weyl tensor term---the Raychaudhuri equation, which gives the evolution of geodesic expansion, is independent of the Weyl tensor. Since the Weyl tensor gives the purely gravitational (i.e., nonlocal) contribution, the singularity in shear-free collapse is essentially nongravitational in nature, being instead determined by the local matter distribution. Indeed, order of magnitude estimates yield $C_{\mu\nu\gamma\delta}\sim R_{\mu\nu}\sim r^{-2}$ for shear-free collapse, whence Weyl fails to dominate over Ricci~\cite{coulomb}.

Finally, a natural question is: {\em How do these results carry over to nonspherical geometries?} The analogous singularity properties of spherical Tolman-Bondi collapse and Szekeres~\cite{szekeres75} (quasispherical) dust collapse~\cite{quasisph} suggest that quasispherical shear-free collapse should remain strongly censored. For stronger departures from spherical symmetry, given the local matter-dominated nature of the singularity, one may speculate that the result holds (see conjecture below), although this is far from evident. We note, however, that Szekeres spacetimes---notwithstanding the fact that they do not admit {\em any} Killing vector field---cannot contain gravitational radiation; for stronger departures from spherical symmetry, there are in general an infinite number of gravitational degrees of freedom, and the singularity might conceivably be Weyl-dominated.

Based on all of the above, we conclude by putting forward the following conjecture: Let $(M,g_{\mu\nu})$ be a spacetime admitting an ADM decomposition $M\approx{\mathbb R}\times\Sigma$, where $\Sigma$ is a submanifold with induced Riemannian three-metric $\gamma_{ij}$ and conjugate gravitational momentum $\pi_{ij}$. Let $T_{\mu\nu}$ be the stress-energy tensor in $(M,g_{\mu\nu})$, and let $u^{\mu}$ be the proper four-velocity of the matter flow, with four-acceleration $a^{\mu}:=u^{\nu}\nabla_{\nu}u^{\mu}$. If the following conditions hold: (i) $(\gamma_{ij},\pi_{ij})$ are regular initial data on $\Sigma$, (ii) $T_{\mu\nu}$ obeys the weak energy condition, and (iii) $\sigma_{\mu\nu}:=\nabla_{(\nu}u_{\mu)}-\fr{1}{3}(\nabla_{\alpha}u^{\alpha})(g_{\mu\nu}-u_{\mu}u_{\nu})+a_{(\mu}u_{\nu)}=0$, then the Cauchy horizon in $(M,g_{\mu\nu})$ is the empty set.

\section*{ACKNOWLEDGMENTS}

I am grateful to Vince Moncrief and Sanjay Jhingan for discussions. This work was supported by FCT Grant SFRH-BPD-5615-2001 and by NSF Grant PHY-0098084.


\begin{thebibliography}{99}
\bibitem{global}
By locally covered, we mean a singularity which is spacelike, or, equivalently, a spacetime whose Cauchy horizon is the empty set. Clearly, all locally covered singularities are also globally covered, but the converse is not true (e.g., a singularity may be locally naked but globally covered, if the Cauchy horizon is non-empty and is a proper subset of the interior of the event horizon).
\bibitem{coulomb}
The Coulomb component of the Weyl tensor in the NP formalism, $\Psi_{(2)}=C_{\alpha\beta\gamma\delta}\delta^{\alpha}_{\theta}\delta^{\beta}_{\varphi}\delta^{\gamma}_{\theta}\delta^{\delta}_{\varphi}$, gives the useful relation $\Psi_{(2)}=\fr{\rho r^{3}}{6R^{3}}-\fr{M}{R^{3}}$. Since $R_{\mu\nu}\sim T_{\mu\nu}\sim\rho\sim r^{-2}$ and $M\sim R\sim r$, it follows that $C_{\alpha\beta\gamma\delta}\sim R_{\mu\nu}\sim r^{-2}$.
\end{thebibliography}
\end{document}